\documentclass[12pt]{article}
\usepackage{graphicx}

\def\a{\alpha}
\def\b{\beta}

\def\d{\delta}

\def\f{\phi}
\def\g{\gamma}

\def\j{\psi}

\def\m{\mu}
\def\n{\nu}

\def\r{\rho}

\def\t{\tau}

\def\D{\Delta}
\def\F{\Phi}

\def\L{\Lambda}

\def\Q{\Theta}

\def\ve{\varepsilon}

\def\tm{\widetilde{\mu}}
\def\bm{\overline{\mu}}

\def\he{\widehat{e}}
\def\be{\overline{e}}
\def\rt{\widetilde{\rho}}
\def\rb{\overline{\rho}}
\def\tM{\widetilde{M}}

\def\beq{\begin{equation}}
\def\eeq{\end{equation}}
\def\bea{\begin{eqnarray}}
\def\eea{\end{eqnarray}}

\def\NO{\nonumber}

\def\pl#1#2#3{Phys.~Lett.~{\bf B {#1}} ({#2}) #3}
\def\np#1#2#3{Nucl.~Phys.~{\bf B {#1}} ({#2}) #3}
\def\prl#1#2#3{Phys.~Rev.~Lett.~{\bf #1} ({#2}) #3}
\def\pr#1#2#3{Phys.~Rev.~{\bf D {#1}} ({#2}) #3}

\def\prep#1#2#3{Phys.~Rep.~{\bf {#1}C} ({#2}) #3}

\begin{document}
\date{\mbox{ }}
\title{
{\normalsize UNIL-IPT-03-3\hfill\mbox{}\\
DESY 03-045\hfill\mbox{}\\
April 2003\hfill\mbox{}}\\
\vspace{2cm} 
\textbf{Quarks and Leptons between\\ Branes and Bulk}\\
[8mm]}
\author{T.~Asaka$^a$, W.~Buchm\"uller$^b$, L.~Covi$^b$\\
\\
{\normalsize \it
$^a$ Institute of Theoretical Physics, 
University of Lausanne, Switzerland}\\
{\normalsize \it
$^b$ Deutsches Elektronen-Synchrotron DESY, Hamburg, Germany}
}
\maketitle

\thispagestyle{empty}

\begin{abstract}
\noindent
We study a supersymmetric SO(10) gauge theory in six dimensions compactified
on an orbifold. Three sequential quark-lepton families are localized at the
three fixpoints where SO(10) is broken to its three GUT subgroups. Split bulk 
multiplets yield the Higgs doublets of the standard model and as additional states 
lepton doublets and down-quark singlets. The physical quarks and leptons are 
mixtures of brane and bulk states. 
The model naturally explains small quark mixings together with 
large lepton mixings in the charged current. A small hierarchy of neutrino masses
is obtained due to the different down-quark and up-quark mass hierarchies.
None of the usual GUT relations between fermion masses holds exactly.
\end{abstract}

\newpage

The explanation of the masses and mixings of quarks and leptons remains a challenge for
theories which go beyond the standard model \cite{fx00,ros01}. In principle, grand
unified theories (GUTs) appear as the natural framework to address this question.
However, as much work on this topic has demonstrated, all simple GUT relations for
fermion mass matrices are badly violated and, within the conventional framework of
four-dimensional (4d) unified theories, a complicated Higgs sector is needed to achieve
consistency with experiment.

In this paper we shall address the flavour problem in the context of a supersymmetric
SO(10) GUT in six dimensions compactified on an orbifold \cite{abc01,hnx02}. A new
ingredient of orbifold GUTs is the presence of split bulk multiplets whose mixings with
complete GUT multiplets can significantly modify ordinary GUT mass relations 
\cite{hn01,hm01}. This extends the well know mechanism of mixing with vectorlike 
multiplets \cite{bar80}. Several analyses of the flavour structure of orbifold GUTs
have already been carried out (cf., e.g., \cite{hks02}-\cite{kr03}). In 5d theories large
bulk mass terms can lead to a localization of zero modes at one of the two
boundary branes, which can explain fermion mass hierarchies \cite{hm02}. In this   
way a realistic `lopsided' structure of Yukawa matrices can be achieved \cite{hmy03}.  

`Lopsided' fermion mass matrices, mostly based on an abelian generation symmetry
\cite{fn79}, have received much attention in recent years (cf.~\cite{sy98}-\cite{af99}).
In the context of SU(5) GUTs they introduce a large mixing of left-handed leptons
and right-handed down quarks, which leads to small mixings among the left-handed
down-quarks. In this way the observed large mixings in the leptonic charged current
can be reconciled with the small CKM mixings in the quark current. The mechanism
of flavour mixing, which we describe below, is also based on large mixings of
left-handed leptons and right-handed down quarks. However, these mixings do not
respect SU(5) and they are not controlled by a single hierarchy parameter. In
this way a different pattern of mixings is achieved with several characteristic
predictions for the neutrino sector. 

Let us now consider SO(10) gauge theory in 6d with $N=1$ supersymmetry compatified on 
the orbifold $T^2/(Z^I_2\times Z^{PS}_2\times Z^{GG}_2)$ \cite{abc01,hnx02}. The 
theory has four fixed points, $O_I$, $O_{GG}$, $O_{fl}$ and $O_{PS}$, located at the
four corners of a `pillow' corresponding to the two compact dimensions (cf.~fig.~1). 
At $O_I$ only supersymmetry is broken whereas SO(10) remains unbroken.
At $O_{GG}$, $O_{fl}$ and $O_{PS}$ SO(10) is broken to its three GUT subgroups 
G$_{GG}$=SU(5)$\times$U(1)$_X$, flipped SU(5),  G$_{fl}$=SU(5)'$\times$U(1)', and 
G$_{PS}$=SU(4)$\times$SU(2)$\times$SU(2), respectively. The intersection of all
these GUT groups yields the standard model group with an additional U(1) factor,
G$_{SM'}$= SU(3)$\times$SU(2)$\times$U(1)$_Y \times$U(1)$_X$, as unbroken gauge symmetry
below the compactification scale. $B-L$, the difference of baryon and lepton number, is
a linear combination of $Y$ and $X$.

The field content of the theory is strongly constrained by the required cancellation of 
irreducible bulk and brane anomalies \cite{abc03}. Motivated by the embedding of all field 
quantum numbers into the adjoint representation of $E_8$ \cite{abc02}, we have  6 
{\bf 10}-plets, $H_1,\ldots, H_6$, and 4 {\bf 16}-plets, $\F, \F^c, \f, \f^c$ as bulk 
hypermultiplets, accompanied by 3 {\bf 16}-plets $\j_i$, $i=1\ldots 3$, of brane fields.
Vacuum expectation values of $\F$ and $\F^c$ break $B-L$. The electroweak gauge group 
is broken by expectation values of $H_1$ and $H_2$.

\begin{figure}
\centering 
\includegraphics[scale=1.2]{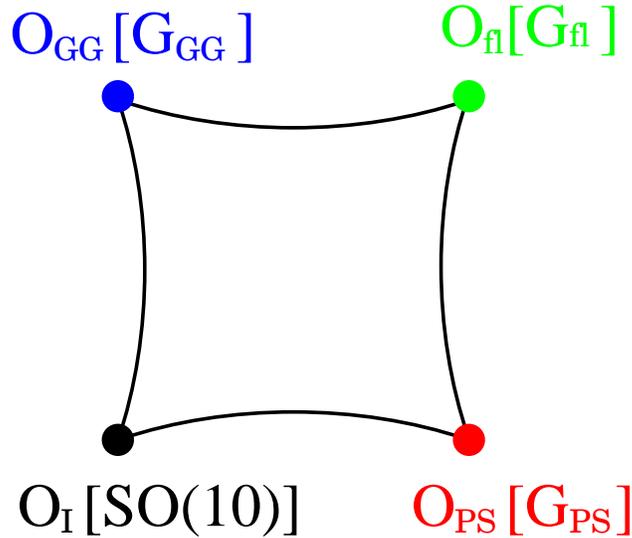}
\caption{The three SO(10) subgroups at the corresponding fixpoints of the
orbifold $T^2/(Z_2^I\times Z_2^{PS}\times Z_2^{GG})$ .\label{fig:orb}}
\end{figure}
  
Compared to \cite{abc02} we have added an additional pair of bulk {\bf 16}-plets,
$\f$ and $\f^c$ together with two {\bf 10}-plets, $H_5$ and $H_6$, to cancel bulk anomalies.
This is still compatible with the embedding in $E_8$, and it corresponds to the largest
number of bulk fields consistent with the cancellation of anomalies. Note that both the
irreducible and reducible 6d gauge anomalies vanish.

The parities of $H_5$, $H_6$ and $\f$ are listed in table~1. $\f^c$ has the same parities 
as $\f$. The corresponding zero modes are
\bea
L = \left(\begin{array}{l} \n_4 \\ e_4 \end{array}\right)\;, \quad  
L^c = \left(\begin{array}{l} \n^c_4 \\ e^c_4 \end{array}\right)\;, \quad  
G^c_5 = d^c_4\;, \quad G_6 = d_4\;.
\eea
The zero modes of the fields $\F$, $\F^c$, $H_1\ldots H_4$ are given in \cite{abc02}.
They are the color triplets and singlets $D^c$, $N^c$, $D$, $N$, $H_1^c$, $H_2$, 
$G^c_3$ and $G_4$.

\renewcommand{\arraystretch}{1.2}
\begin{table}[t]
\begin{center}
   $\begin{array}[h]{|c||cc|cc|cc|cc|}\hline
     \mbox{SO(10)} &
     \multicolumn{8}{|c|}{ \mathbf{10} }
     \\ \hline
     G_{PS} &
     \multicolumn{2}{|c|}{ ( \mathbf{1}, \mathbf{2}, \mathbf{2}) } &
     \multicolumn{2}{|c|}{ ( \mathbf{1}, \mathbf{2}, \mathbf{2}) } &
     \multicolumn{2}{|c|}{ ( \mathbf{6}, \mathbf{1}, \mathbf{1}) } &
     \multicolumn{2}{|c|}{ ( \mathbf{6}, \mathbf{1}, \mathbf{1}) }
     \\ \hline
     G_{GG} &
     \multicolumn{2}{|c|}{ \mathbf{5}^\ast{}_{-2} } &
     \multicolumn{2}{|c|}{ \mathbf{5}{}_{+2} } &
     \multicolumn{2}{|c|}{ \mathbf{5}^\ast{}_{-2} }  &
     \multicolumn{2}{|c|}{ \mathbf{5}{}_{+2} }
     \\ \hline
        &  \multicolumn{2}{|c|}{H^c} & \multicolumn{2}{|c|}{H} &
     \multicolumn{2}{|c|}{G^c} & \multicolumn{2}{|c|}{G}
     \\
     {} &
    Z_2^{PS} & Z_2^{GG} &
    Z_2^{PS} & Z_2^{GG} &
    Z_2^{PS} & Z_2^{GG} &
    Z_2^{PS} & Z_2^{GG}
    \\ \hline \hline
     H_5 &
     - & + &
     - & - &
     + & + &
     + & -
     \\ \hline \hline
     H_6 &
     - & - &
     - & + &
     + & - &
     + & + 
    \\ \hline
    \end{array}$
  \end{center}
  \begin{center}
  $\begin{array}[h]{|c||cc|cc|cc|cc|}\hline
    \mbox{SO(10)} & \multicolumn{8}{|c|}{ \mathbf{16} }
    \\ \hline
    G_{PS} &
    \multicolumn{2}{|c|}{ (\mathbf{4}, \mathbf{2}, \mathbf{1}) } &
    \multicolumn{2}{|c|}{ (\mathbf{4}, \mathbf{2}, \mathbf{1}) } &
    \multicolumn{2}{|c|}{ (\mathbf{4}^\ast, \mathbf{1}, \mathbf{2}) } &
    \multicolumn{2}{|c|}{ (\mathbf{4}^\ast, \mathbf{1}, \mathbf{2}) }
    \\ \hline
    G_{GG} &
    \multicolumn{2}{|c|}{ \mathbf{10}_{-1} } &
    \multicolumn{2}{|c|}{ \mathbf{5}^\ast{}_{+3} } &
    \multicolumn{2}{|c|}{ \mathbf{10}_{-1 }} &
    \multicolumn{2}{|c|}{ \mathbf{5}^\ast{}_{ +3 }, \mathbf{1}_{ -5 } }
    \\ \hline
    {} &
    \multicolumn{2}{|c|}{ Q}      &
    \multicolumn{2}{|c|}{L}       &
    \multicolumn{2}{|c|}{U, E}  &
    \multicolumn{2}{|c|}{D^c, N^c}
    \\
    {} &
    Z_2^{PS} & Z_2^{GG} &
    Z_2^{PS} & Z_2^{GG} &
    Z_2^{PS} & Z_2^{GG} &
    Z_2^{PS} & Z_2^{GG}
     \\ \hline
     \phi &
     + & - &
     + & + &
     - & - &
     - & +
     \\ \hline
    \end{array}$
    \caption{Parity assignments for the bulk hypermultiplets $H_5$, $H_6$ and $\f$.}
    \label{tab:P16}
  \end{center}
\end{table}

Fermion masses and mixings are determined by brane superpotentials. The allowed terms
are restricted by R-invariance and an additional U(1)$_{\tilde{X}}$ symmetry \cite{abc02}. 
The corresponding charges of the superfields are given in table~\ref{tab:rpa}. The fields
$H_1$, $H_2$, $\F$ and $\F^c$, which aquire a vacuum expectation value, have vanishing
R-charge. All matter fields have R-charge one. Since $\j_i$ and $\f$ have the same charges
we combine them to the quartet $(\j_\a) = (\j_i,\f)$, $\a= 1\ldots4$. The most general 
brane superpotential up to quartic terms is then given by
\bea\label{O10}
W &=&  M^d H_5 H_6 + M^l_\a \j_\a \f^c + M_{12} H_1 H_3 + M_{23} H_2 H_3 \NO\\ 
&& + {1\over 2} h^{(1)}_{\a\b} \j_\a \j_\b H_1 + {1\over 2} h^{(2)}_{\a\b} \j_\a \j_\b H_2 
+ f_\a \F \j_\a H_6 + f_5 \F^c \f^c H_5 \NO\\
&& + f^D \F^c \F^c H_3 + f^G \F \F H_4 + 
{1\over 2}{h^N_{\a\b}\over M_*}\j_\a\j_\b\F^c\F^c \NO\\
&& + {k_1\over M_*} H_1^2 H_5^2 + {k_2\over M_*} H_1 H_2 H_5^2 + {k_3\over M_*} H_2^2 H_5^2
+ {k_4 \over M_*} \F \F^c H_1 H_3 \NO\\
&& + {k_5 \over M_*} \F \F^c H_2 H_3 + {g^d_\a \over M_*} \F^c \j_\a H_5 H_1 
+ {g^u_\a \over M_*} \F^c \j_\a H_5 H_2 + {g^d \over M_*} \F \f^c H_5 H_1 \NO\\
&&+ {g^u \over M_*} \F \f^c H_5 H_2 + {k^d_\a \over M_*} \F \F^c \j_\a \f^c 
+ {k^l_\a \over M_*} \F \F^c  \j_\a \f^c + {k^l \over M_*} \F \F \f^c \f^c \;, 
\eea
where we choose $M_* > 1/R_{5,6} \sim \L_{GUT}$ to be the cutoff of the 6d theory, and the
bulk fields have been properly normalized. All the volume factors due to the 6d fields are 
absorbed into the unknown couplings and we will not use them to explain the hierarchies.
When the bulk fields are replaced by their
zero modes only 9 of the 23 terms appearing in the superpotential remain. Although we
have written the superpotential in terms of SO(10) multiplets, on the different
branes the Yukawa couplings $h^{(1)}$ and $h^{(2)}$ split into $h^{(d)}, h^{(e)}$ and 
$h^{(u)}, h^{(D)}$, respectively. Some of these couplings are equal due to GUT relations
on the corresponding brane.

\renewcommand{\arraystretch}{1.2}
\begin{table}[t]
  \begin{center}
    $\begin{array}[h]{| c | c c c c c c c c c c c |}
      \hline
      {} & H_1 & H_2 & \F^c & H_3 & \F & H_4 & \j_i & \f^c & \f & H_5 & H_6 
      \\ \hline
      R &  0 & 0 & 0 & 2 & 0 & 2 & 1 & 1 & 1 & 1 & 1
   \\ \hline
      \widetilde{X} & - 2a & - 2a & - a & 2a & a & - 2a & a & - a & a & 2a & -2a   
   \\ \hline
    \end{array}$
    \caption{Charge assignments for the symmetries U(1)$_R$ and U(1)$_{\tilde{X}}$.}
    \label{tab:rpa}
  \end{center}
\end{table}

The main idea to generate fermion mass matrices is now as follows. We consider the case
that the three sequential {\bf 16}-plets are located on the three branes where SO(10)
is broken to its three GUT subgroups. As an example, we place $\j_1$ at $O_{GG}$, 
$\j_2$ at $O_{fl}$ and $\j_3$ at $O_{PS}$. The three `families' are then separated by 
distances large compared to the cutoff scale $M_*$. Hence, they can only have diagonal 
Yukawa couplings with the bulk Higgs fields. Direct mixings are exponentially suppressed. 
However, the brane fields can mix with the bulk zero modes for which we expect no 
suppression. These mixings take place only among left-handed leptons and right-handed 
down-quarks. This leads to a characteristic pattern of mass matrices which we shall now 
explore.

If $B-L$ is broken, as discussed in \cite{abc02}, 
$\langle\F^c\rangle = \langle\F\rangle = v_N$, and the bulk zero modes $N^c$, $N$,
($D,G^c$) and ($D^c,G$) aquire masses ${\cal O}(v_N)$. After electroweak symmetry breaking,
with $\langle H^c_1 \rangle = v_1$, $\langle H_2 \rangle = v_2$,
the remaining states have the following mass terms, 
\bea
W &=& d_\a m^d_{\a\b} d^c_\b + e^c_\a m^e_{\a\b} e_\b + n^c_\a m^D_{\a\b} \n_\b \NO\\
&& + u^c_i m^u_{ij} u_j + {1\over 2} n^c_i M_{ij} n^c_j\;.
\eea
Here $m^d$, $m^e$ and $m^D$ are $4\times 4$ matrices,
\bea\label{md}
m^d = \left(\begin{array}{cccc}
h^d_{11}v_1 & 0 & 0 & g_1^d {v_N\over M_*} v_1 \\
0 & h^d_{22}v_1 & 0 & g_2^d {v_N\over M_*} v_1 \\
0 & 0 & h^d_{33}v_1 & g_3^d {v_N\over M_*} v_1 \\
f_1 v_N & f_2 v_N & f_3 v_N & M^d \end{array}\right)\;,
\eea
\bea\label{me}
m^e = \left(\begin{array}{cccc}
h^d_{11}v_1 & 0 & 0 & h_{14}^e v_1 \\
0 & h^e_{22}v_1 & 0 & h_{24}^e v_1 \\
0 & 0 & h^d_{33}v_1 & h_{34}^e v_1 \\
M^l_1 & M^l_2 & M^l_3 & M^l_4 \end{array}\right)\;,
\eea

\bea\label{mD}
m^D = \left(\begin{array}{cccc}
h^D_{11}v_2 & 0 & 0 & h_{14}^D v_2 \\
0 & h^u_{22}v_2 & 0 & h_{24}^D v_2 \\
0 & 0 & h^u_{33}v_2 & h_{34}^D v_2 \\
M^l_1 & M^l_2 & M^l_3 & M^l_4 \end{array}\right)\;,
\eea
whereas $m^u$ and $m^N$ are diagonal $3\times 3$ matrices,
\bea\label{muN}
m^u = \left(\begin{array}{ccc}
h^u_{11}v_2 & 0 & 0 \\
0 & h^u_{22}v_2 & 0 \\
0 & 0 & h^u_{33}v_2 \end{array}\right)\;, \quad
m^N = \left(\begin{array}{ccc}
h^N_{11}{v_N^2\over M_*} & 0 & 0 \\
0 & h^N_{22}{v_N^2\over M_*} & 0 \\
0 & 0 & h^N_{33}{v_N^2\over M_*} \end{array}\right)\;.
\eea
In the matrices $m^d$, $m^e$ and $m^D$ we have neglected corrections ${\cal O}(v_N/M_*)$.
The diagonal elements satisfy four GUT relations which correspond to the unbroken SU(5), 
flipped SU(5) and Pati-Salam subgroups of SO(10).

The crucial feature of the matrices $m^d$, $m^e$ and $m^D$ are the mixings between
the six brane states and the two bulk states. The first three rows of the matrices
are proportional to the electroweak scale. The corresponding Yukawa couplings have
to be hierarchical in order to obtain a realistic spectrum of quark and lepton masses.
This corresponds to different strengths of the Yukawa couplings at the different
fixpoints of the orbifold. The fourth row, proportional to $M^d$, $M^l$ and $v_N$, is 
of order the unification scale and, we assume, non-hierarchical.

The mass matrices $m^d$, $m^e$ and $m^D$ are of the form
\bea
m = \left(\begin{array}{cccc}
\m_1 & 0 & 0 & \tm_1 \\
0 & \m_2 & 0 & \tm_2 \\
0 & 0 & \m_3 & \tm_3 \\
\tM_1 & \tM_2 & \tM_3 & \tM_4 \end{array}\right)\;,
\eea
where $\m_i,\tm_i = {\cal O}(v_{1,2})$ and $\tM_i = {\cal O}(\L_{GUT})$. To diagonalize 
the matrix $m$ it is convenient to define a set of four-dimensional unit vectors as follows,
\bea
(\tM_1,\ldots \tM_4) = \tM e_4^T\;, \quad 
e^T_\a e_\b = e^T_{\a\g} e_{\b\g} = \d_{\a\b}\;.
\eea
Using the orthogonal matrices ($\a,\b = 1\ldots 4$, $i = 1 \ldots 3$),
\bea
V_{\a\b} = (e_\b)_\a \;, \quad 
U_{\a\b} = \d_{\a\b} - {1 \over \tM} \d_{\a 4}(e_{4i}\m_i 
+ e_{44}\tm_i)\d_{\b i} + {\cal O}\left({v^2 \over \tM^2}\right) \;,
\eea 
we can now perform a change of basis which yields for the mass matrix, 
\bea
m' = U^T m V = \left(\begin{array}{cc}
\widehat{m} & 0 \\ 0 & \tM \end{array}\right) + {\cal O}\left({v^2\over \tM^2}\right)\;,
\eea
where the $3\times 3$ matrix $\widehat{m}$ is given by
\bea\label{mhat}
\widehat{m} = \left(\begin{array}{c}
\m_1 \he^T_1 + \tm_1 \he^T_4  \\ 
\m_2 \he^T_2 + \tm_2 \he^T_4  \\ 
\m_3 \he^T_3 + \tm_3 \he^T_4   \end{array}\right)\;.
\eea
Here the three-vectors $\he_\a$, $\a = 1\ldots 4$, are determined by the
four-vectors $e_i$, $i = 1\ldots 3$, with $(\he_\a)_i = (e_i)_\a$.
Note that $\widehat{m}$ is composed of three row vectors of hierarchical length,
a structure familiar from lopsided fermion mass models.

The hierarchy of the row vectors suggests to perform a further change of basis
such that all remaining mixings are small. Three orthogonal three-vectors $\be_i$,
$\be^T_i\be_j = \be_{ik}\be_{jk} = \d_{ij}$, can be defined by writing the matrix
$\widehat{m}$ in the following form
\bea\label{mhat2}
\widehat{m} = \left(\begin{array}{l}
\bm_1 (\g\be^T_1 + \be^T_2 + \b\be^T_3)  \\ 
\bm_2 (\be^T_2 + \a\be^T_3)  \\ 
\bm_3 \be^T_3  \end{array}\right)\;.
\eea
The parameters $\bm_i$ are ${\cal O}(\m_i,\tm_i)$ and therefore again hierarchical.
With respect to this new basis the matrix $m$ has triangular form,
\bea
\overline{m} = \left(\begin{array}{lll}
\bm_1\g & \bm_1 & \bm_1\b \\
0 & \bm_2 & \bm_2\a \\
0 & 0 & \bm_3 \end{array}\right)\;.
\eea
For our discussion of mass eigenvalues and mixing angles we shall need the two
matrices $m m^T$ and $m^T m$, which in the basis $\be_i$ are both hierarchical,
\bea\label{m2t}
m m^T = \left(\begin{array}{lll}
\bm_1^2 (1 + \b^2 + \g^2) & \bm_1\bm_2 (1 + \a\b) & \bm_1\bm_3 \b \\
\bm_1\bm_2 (1 + \a\b) & \bm_2^2 (1 + \a^2) & \bm_2\bm_3 \a \\
\bm_1\bm_3 \b & \bm_2\bm_3 \a & \bm_3^2 \end{array}\right)\;,
\eea
\bea\label{mt2}
m^T m = \left(\begin{array}{lll}
\bm_1^2 \g^2 & \bm_1^2 \g & \bm_1^2 \b\g  \\
\bm_1^2 \g & \bm_2^2 + \bm_1^2 & \bm_2^2 \a + \bm_1^2 \b \\
\bm_1^2 \b\g &  \bm_2^2 \a + \bm_1^2 \b & \bm_3^2 + \bm_2^2 \a^2 + \bm_1^2 \b^2
\end{array}\right)\;.
\eea

Consider now the up-quark mass matrix. We concentrate on the case of large 
$\tan{\b} = v_2/v_1 \simeq 50$, such that $h^d_{33} \simeq h^u_{33}$. The diagonal 
elements of the mass matrices (\ref{md}), (\ref{me}), (\ref{mD}) and (\ref{muN}) 
are partially connected by the GUT relations which hold on the different branes.
For simplicity, we therefore assume universally,
\bea\label{diag}
\m_1 : \m_2 : \m_3 \sim m_u : m_c : m_t \;.
\eea
It is well known that the hierarchy of down-quark and charged lepton masses is
substantially smaller than the up-quark mass hierarchy. Given the scaling (\ref{diag})
of the diagonal elements and the structure of $m^d$ and $m^e$ this 
implies that the down-quark and charged lepton mass matrices must be dominated by
the off-diagonal elements. Hence, we assume again universally,
\bea
\m_1 \ll \bm_1 \sim \tm_1\;, \quad  \m_2 \ll \bm_2 \sim \tm_2\;, 
\quad \m_3 \sim \bm_3\;.
\eea 
The parameters $\bm_{1,2}$ of the matrix $\overline{m}$ are then dominated by the mixing terms
$\tm_{1,2}$, i.e. $\bm_{1,2} \sim \tm_{1,2}$.

Since the up-quark matrix $m^u$ is diagonal the CKM quark mixing matrix is given
by the matrix $V$ which diagonalizes $m^d m^{dT}$. From eq.~(\ref{m2t}) one reads
off for the two larger masses 
\bea
m_b \simeq \bm_3 \;, \quad m_s \simeq \tm_2 \;, 
\eea
and for the mixing angles
\bea
V_{us}=\Q_c \sim {\tm_1\over \tm_2}\;, \quad
V_{cb} \sim {\tm_2\over \tm_3}\;, \quad
V_{ub} \sim {\tm_1\over \tm_3}\;.
\eea
Using $m_b$, $m_s$ and $\Q_c \simeq 0.2$ as input one obtains for the two remaining
mixing angles
\bea
V_{cb} \sim {m_s\over m_b} \simeq 2\times 10^{-2} \;,\quad
V_{ub} \sim \Q_c {m_s\over m_b} \simeq 4\times 10^{-3} \;,
\eea
in agreement with analyses of weak decays \cite{fle02} up to a factor of two, which is 
beyond the predictivity of our approach. 

The smallest eigenvalue vanishes in the limit $\m_1,\m_2 \rightarrow 0$, since in
this case two vectors of the matrix $\widehat{m}$ become parallel, with
$\b = \a$ and $\g = 0$. Choosing, for simplicity, $\m_1/\tm_1 < \m_2/\tm_2$, one
has for non-zero $\m_1, \m_2$,
\bea\label{gamma}
\g \sim {\m_2 \over \tm_2} \sim {m_c m_b\over m_t m_s} \sim 0.1 \;.
\eea
This relation will also be important in our analysis of the neutrino masses. For
the down-quark mass one obtains
\bea
{m_d\over m_s} \sim {\m_2\over \tm_2}{\tm_1\over \tm_2}
\sim \ \Q_c \; {m_c m_b\over m_t m_s}  \simeq\ 0.03\;,
\eea
consistent with data \cite{fx00}.

The charged lepton mass matrix $m^e$ is very similar to the down-quark mass matrix.
The main difference is that now there are large mixings between the `left-handed'
states $e_i$. To obtain the contribution of the charged leptons to the leptonic
mixing matrix we consider the matrix $m^{eT}m^e$ as given in eq.~(\ref{mt2})
in the basis $\be_i$. For the two large eigenvalues of $m^e$ one has
$m_{\t} \sim \bm_3 \sim m_b$ and $m_{\m} \sim \bm_2 \sim m_s$. These relations are
consistent with data within our accuracy. A potential problem is the smallness of
the electron mass, i.e. $m_e/m_\m \simeq 0.1\; m_d/m_s$. The smallest eigenvalue of
$m^e$ is again given by $m_e/m_\m \sim (\m_2 \tm_1/ \tm_2^2)$.
However,  in our model the usual SU(5) relations don't hold for the second row
of the mass matrices. Hence, the electron mass is not determined by down quark masses.

Using the diagonal and off-diagonal elements of the mass matrices as determined from
up- and down-quark mass matrices, we can now discuss the implications for neutrino
masses. The heavy Majorana neutrinos scale like up-quarks (cf.~(\ref{muN})),
\bea
M_3 : M_2 : M_1 \sim m_t : m_c : m_u \;.
\eea
The light neutrino masses are given by the seesaw relation
\bea\label{seesaw}
m_\n = - m^{DT} {1\over M^N} m^D\;.
\eea
The structure of the charged lepton and the Dirac neutrino mass matrices 
(cf.~(\ref{me}),(\ref{mD})) is the same. Both matrices lead to large mixings
between the `left-handed' states. In order to determine the leptonic mixing
matrix we discuss the Dirac neutrino matrix in the basis $\be_i$ where the 
remaining mixings of the left-handed charged leptons is small by construction
(cf.~(\ref{mt2})). 

The Dirac neutrino mass matrix can be written as (cf.~(\ref{mhat})),
\bea
\widehat{m}^D = \left(\begin{array}{c}
\r_1 \he^T_1 + \rt_1 \he^T_4  \\ 
\r_2 \he^T_2 + \rt_2 \he^T_4  \\ 
\r_3 \he^T_3 + \rt_3 \he^T_4   \end{array}\right)\;.
\eea
Here the parameters $\r_i$, $\rt_i$ are expected to have the same hierarchy as
$\m_i$, $\tm_i$. However, in general these parameters will differ by factors 
${\cal O}(1)$ since there the entries of $m^e$ and $m^D$ arise from different 
Yukawa couplings in the superpotential. This implies for the matrix 
$\widehat{m}^D$,
with respect to the vectors $\be_i$,
\bea
\widehat{m}^D = \left(\begin{array}{l}
\rb_1 (A\be^T_1 + D\be^T_2 + \be^T_3)  \\ 
\rb_2 (B\be^T_1 + E\be^T_2 + \be^T_3) \\ 
\rb_3 (C\be^T_1 + F\be^T_2 + \be^T_3) \end{array}\right)\;,
\eea
where $\rb_i \simeq \rt_i$.
Hence, with respect to the basis $\bm_i$ the matrix $\widehat{m}^D$ has no longer triangular
form,
\bea\label{mbn}
\overline{m}^D = \left(\begin{array}{lll}
A\rb_1 & B\rb_1 & \rb_1 \\
C\rb_2 & D\rb_2 & \rb_2 \\
E\rb_3 & F\rb_3 & \rb_3 \end{array}\right)\;.
\eea
Generically, the parameters $A \ldots F$ are all ${\cal O}(1)$. All we know is that
for $\mu_{1,2}=\r_{1,2} = 0$ the first two row vectors are parallel, with $A=B=C=0$ and
$D=E$. For $\mu_{1,2},\r_{1,2} \neq 0$ one has analogous to the charged lepton mass matrix 
(cf.~(\ref{gamma})),
\bea\label{esti}
A, B, C, D-E \sim {\r_2\over \rt_2} \sim {\m_2 \over \tm_2} \sim \g \sim 0.1\;.
\eea 

From eqs.~(\ref{seesaw}) and (\ref{mbn}) one now obtains for the light neutrino mass matrix,
\bea\label{mln}
-\overline{m}_\n = \overline{m}^{DT}{1\over M^N}\overline{m}^D = \hspace{11cm} \\
\mbox{} \NO\\
\left(\begin{array}{lll}
A^2{\rb_1^2\over M_1}+B^2{\rb_2^2\over M_2}+C^2{\rb_3^2\over M_3} &
AD{\rb_1^2\over M_1}+BE{\rb_2^2\over M_2}+CF{\rb_3^2\over M_3} &
A{\rb_1^2\over M_1}+B{\rb_2^2\over M_2}+C{\rb_3^2\over M_3} \\
AD{\rb_1^2\over M_1}+BE{\rb_2^2\over M_2}+CF{\rb_3^2\over M_3} &
D^2{\rb_1^2\over M_1}+E^2{\rb_2^2\over M_2}+F^2{\rb_3^2\over M_3} &
D{\rb_1^2\over M_1}+E{\rb_2^2\over M_2}+F{\rb_3^2\over M_3} \\
A{\rb_1^2\over M_1}+B{\rb_2^2\over M_2}+C{\rb_3^2\over M_3} &
D{\rb_1^2\over M_1}+E{\rb_2^2\over M_2}+F{\rb_3^2\over M_3} &
{\rb_1^2\over M_1}+{\rb_2^2\over M_2}+{\rb_3^2\over M_3} 
\NO\end{array}\right)\;.
\eea
Using eq.~(\ref{esti}) one immediately sees the order of magnitude of the different 
entries,
\bea
\overline{m}_\n \sim \left(\begin{array}{lll}
\g^2 & \g & \g \\
\g & 1 & 1 \\
\g & 1 & 1 \end{array}\right)m_3\;,
\eea
where $m_3$ is the largest neutrino mass, i.e. $m_1\leq m_2 \leq m_3$.
It is well known that such a matrix can account for all neutrino data. It has 
previously been derived based on a U(1) family symmetry \cite{sy98,ilr98} and
also by requiring a compensation between the Dirac and Majorana neutrino mass 
hierarchies \cite{bw01,xin02}.

Consider now the parameters in the matrix (\ref{mln}). The mass matrices $m^d$,
$m^e$ and $m^D$ have the same structure with large off-diagonal entries. For simplicity,
we therefore assume for the mass parameters $\bm_i$ and $\rb_i$ have a similar hierarchy,
approximately given by the down-quark masses, i.e.
$\rb_1 : \rb_2 : \rb_3 \sim m_d : m_s : m_b$. One then obtains
\bea\label{seq}
{\rb_2^2\over M_2}{M_3\over \rb_3^2} \sim {m_s^2 m_t\over m_b^2 m_c} \sim 0.2 \;, \quad
{\rb_1^2\over M_1}{M_3\over \rb_3^2} \sim {m_d^2 m_t\over m_b^2 m_u} \sim 0.2 \;.
\eea 
This corresponds to the picture of sequential heavy neutrino dominance \cite{kin00}.
It yields large 2-3 mixing, $\sin{2\Q_{23}}\sim 1$. The largest neutrino mass is
$m_3 \sim m_t^2/ M_3$.
Identifying $m_3$ with $\sqrt{\D m_{atm}^2} \sim 0.05$~eV one obtains for the heavy
Majorana masses $M_3 \sim 10^{15}$~GeV, $M_2 \sim 3\times 10^{12}$~GeV and
$M_1 \sim  10^{10}$~GeV. The second neutrino mass is $m_2 \sim 0.01$~eV,
which is consistent with data within our accuracy.

Since the 2-3 determinant is small the matrix (\ref{mln}) can also account for the
LMA MSW-solution of the solar neutrino problem \cite{vis98}. As all neutrino 
masses are rather close to each other, with unknown coefficients ${\cal O}(1)$, a 
precise prediction of the mixing angle $\Q_{12}$ and the smallest neutrino mass is 
not possible. Generically, one has $\sin{2\Q_{12}} \sim \g m_3/m_2$ and
$m_1 = {\cal O}(\g m_3, m_2)$. On the other hand, a definitive prediction of the matrix 
(\ref{mln}) is a rather large 1-3 mixing angle, $\Q_{13} \sim \g \sim 0.1$.

Decays of the lightest right-handed neutrinos may be the origin of the baryon asymmetry of
the universe \cite{fy86}. In addition to the mass $M_1 \sim 10^{10}$~GeV the relevant
quantities are the CP-asymmetry $\ve_1$ and the effective neutrino mass 
$\widetilde{m} = (m^{D\dagger}m^D)_{11}/M_1$. One easily obtains 
$\ve_1 \sim 0.1\; M_1/M_3 \sim 10^{-6}$ and $\widetilde{m}_1 \sim 0.2\; m_3$. 
These are the typical parameters of thermal leptogenesis \cite{bp96}.

Starting from three sequential families located at three different fixpoints of an orbifold,
we have shown that the mixing with split bulk multiplets can lead to a characteristic
pattern of quark and lepton mass matrices which can account for small quark mixings
together with large lepton mixings in the charged current. Correspondingly, the quark mass
hierarchies are large whereas the small neutrino mass hierarchy follows from the difference
of down-quark and up-quark mass hierarchies. The dynamical origin of the hierarchy of Yukawa
couplings at the different branes remains to be understood.

\noindent
We would like to thank A.~Hebecker and D.~Wyler for helpful discussions.


\end{document}